# Surface magnetism in $ZnO/Co_3O_4$ mixtures

M.A. Garcia[1,2,*], F. Jiménez-Villacorta[3], A.Quesada[4], J. de la Venta[1,2], N. Carmona[2], I. Lorite[1], J. Llopis[2], and J. F. Fernández[1].

1 Instituto de Cerámica y Vidrio, CSIC, Madrid, Spain
2 Dpto. Física de Materiales Universidad Complutense de Madrid, Madrid, Spain
3 SpLine, Spanish CRG beamline at the ESRF, Grenoble, France and Instituto de Ciencia de Materiales, CSIC, Madrid,Spain.
4 Lawrence Berkeley National Laboratory, Berkeley, California 94720, USA

## Abstract

We recently reported the observation of room temperature ferromagnetism in mixtures of ZnO and $Co_3O_4$ despite the diamagnetic and antiferromagnetic character of these oxides respectively. Here we present a detailed study on the electronic structure of this material in order to account for this unexpected ferromagnetism. Electrostatic interactions between both oxides lead to a dispersion of $Co_3O_4$ particles over the surface of ZnO larger ones. As a consequence, the reduction of $Co^{+3}$ to $Co^{2+}$ at the particle surface takes place as evidenced by XAS measurements and optical spectrocopy. This reduction allows to xplain the observed ferromagnetic signal within the well established theories of magnetism.

## 1. Introduction

In the last years, a large number of reports shown unexpected ferromagnetism in oxides, mainly in thin films and nanostructured materials (for a general review see [1]). These effects have been ascribed to diluted magnetic semiconductor oxides (DMSO) [2,3,4] , doping induced defects [5,6,7], nanoscale, surface [8] and proximity effects [9]. However, there is still a large controversy about these new results including possible experimental errors [10], and the origins of this phenomenology are not clear at all. We recently reported room temperature (RT) ferromagnetism (FM) in mixtures of ZnO and $Co_3O_4$ despite the diamagnetic and antiferromagnetic character of both oxides respectively [11]. As figure 1 summarises, the FM appears after soft milling of the mixed oxides and decreases after annealing, disappearing for thermal treatments over 700ºC. Therefore, the effect arises when surfaces of both oxides become in contact and weakly interact and disappear as both oxides react to form a complex oxide (discarding the possibility to have a DMSO). At that time we could not propose an explanation for the appearing of the RT FM but just indicate that the effect exists, it was not a measuring artefact or due to metallic Co segregation nor sample contamination. We present here a study of the electronic structure of the ZnO/$Co_3O_4$ mixtures in order to understand the origin of surprising effect. We show that the presence of RT FM is a surface effect that can be explained with the well known theories of the magnetism of oxides.

## 2. Experimental

$ZnO_1$-x$(Co_3O_4)_X$ samples with x=0.01, 0.05 and 0.25 were prepared as already described in Ref. [11]. Briefly, powders were attrition milled in water medium with zirconia balls, dried, sieved, and pre-reacted afterwards at 400 ºC for 8 h separately in an alumina crucible. Calcined powders were mixed and attrition milled again, and the dried and sieved powders pressed into disks of 20 mm in diameter and 2 mm in thickness that where annealed at different temperatures ranging from 500ºC to 1000ºC in air. Two set of samples were prepared from high purity raw powders independently in two different laboratories (filiations 1&2) and using materials from different suppliers.

The structural analysis of the samples was carried out with a Siemens D5000 X-Ray Diffractometer using a monochromatic Cu K line and operating at 40 kV and 40 mA. The microstructure of the samples was observed with a Field Emission Scanning Electron Microscope, FE-SEM, (Hitachi S-4700, Japan) coupled with Energy Dispersed Spectroscopy, EDS. Magnetic characterization was performed in two different Vibrating Sample Magnetometers: a VSM LDJ Instruments, and a VSM Lakeshore 7304. For the magnetic measurements, all possible sources of experimental errors described in [10] were taken into account. Optical absorption was measured with a Shimadzu 3101 spectrophotometer attached with an integrating sphere. X-ray absorption spectroscopy measurements at the Co K-edge energies were performed in the transmission mode at the Spanish CRG beamline (SpLine), at the ESRF. Two gas ionization chambers, filled with nitrogen and argon, were used to measure the incident and the transmitted beam, respectively. Several scans were taken, in order to obtain a good signal-to-noise ratio.

We will focus here on the samples consisting on 5%$Co_3O_4$-95%ZnO; For the samples with 1%$Co_3O_4$, the X-ray absorption spectroscopy and magnetic measurements were noisy and close to the resolution limit of the equipment (although magnetic measurements shown the highest values when normalised to the $Co_3O_4$ content). On the contrary, for the samples with 25% of $Co_3O_4$ the effect is quite reduced, probably because a smaller fraction of $Co_3O_4$ is in contact with ZnO.

## 3. Results and Discussion

Figure 1 summarises the XRD and magnetic measurements already reported [11]. Briefly, the XRD patterns (figure 1a) show just the peaks corresponding to ZnO and $Co_3O_4$ for the samples milled and annealed up to 1000ºC. Diffractograms of the samples milled and annealed at 500ºC and 600ºC results equivalent indicating the presence of both oxides; annealing at higher temperatures yield a decrease of the $Co_3O_4$ peaks without any other new appearing in the diffraction pattern. This result indicates that there is no significant diffusion up to 600ºC while at larger temperatures the Co diffuses into ZnO to form a $Zn_{1-x}Co_xO$ solid solution, isostructural with the ZnO (and therefore explaining the absence of new peaks in the XRD patterns).

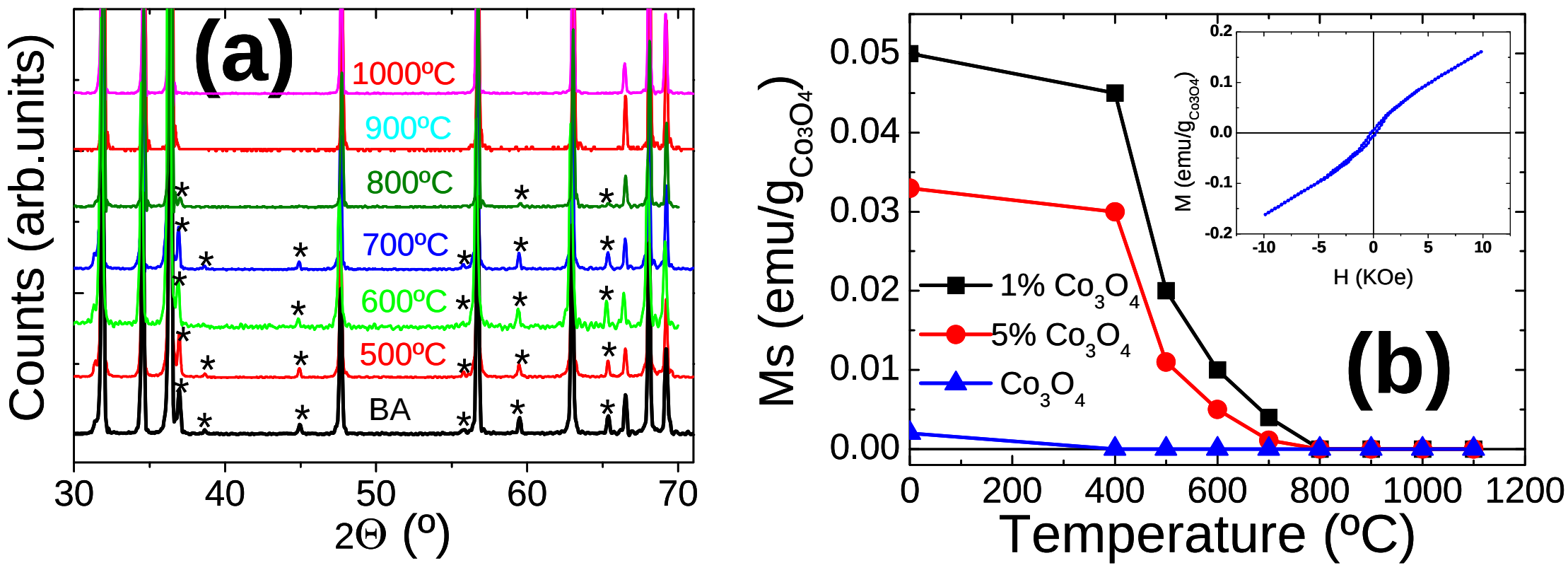

Figure 1 (a) X-ray diffraction patterns from the samples with 95%ZnO-5%Co3O4 . Asterisks indicate $Co_3O_4$ peaks while non-indexed peaks correspond to ZnO. BA stands for Before Annealing (b) Saturation magnetization as a function of the annealing temperature. Inset shows the magnetization curve at 300 K for the sample with 5% $Co_3O_4$ milled.

Samples exhibited paramagnetic behaviour at RT with a superimposed ferromagnetic contribution that disappears after annealing at 700 ºC (figure 1b). The magnetic effects were found to be reproducible for both sets of samples (i.e., the observation of room temperature ferromagnetic component for samples annealed up to 600ºC), but changes in the magnetization values up to 35% were found between both sets (figure 1b shows the data for the set exhibiting the largest values in $M_S$).

## 3.1 SEM analysis

Figure 2 show a FE-SEM image of a sample consisting on 5%$Co_3O_4$-95%ZnO softly milled. After milling, $Co_3O_4$ particles resulted deagglomerated and dispersed over the surface of large ZnO particles. This dispersion is the fingerprint of a strong interaction between both oxides; the reason for this effect is related to differences of surface charge between ZnO and $Co_3O_4$ that promotes the dispersion of $Co_3O_4$ onto ZnO large particle driven by electrostatic forces [12,13 ].

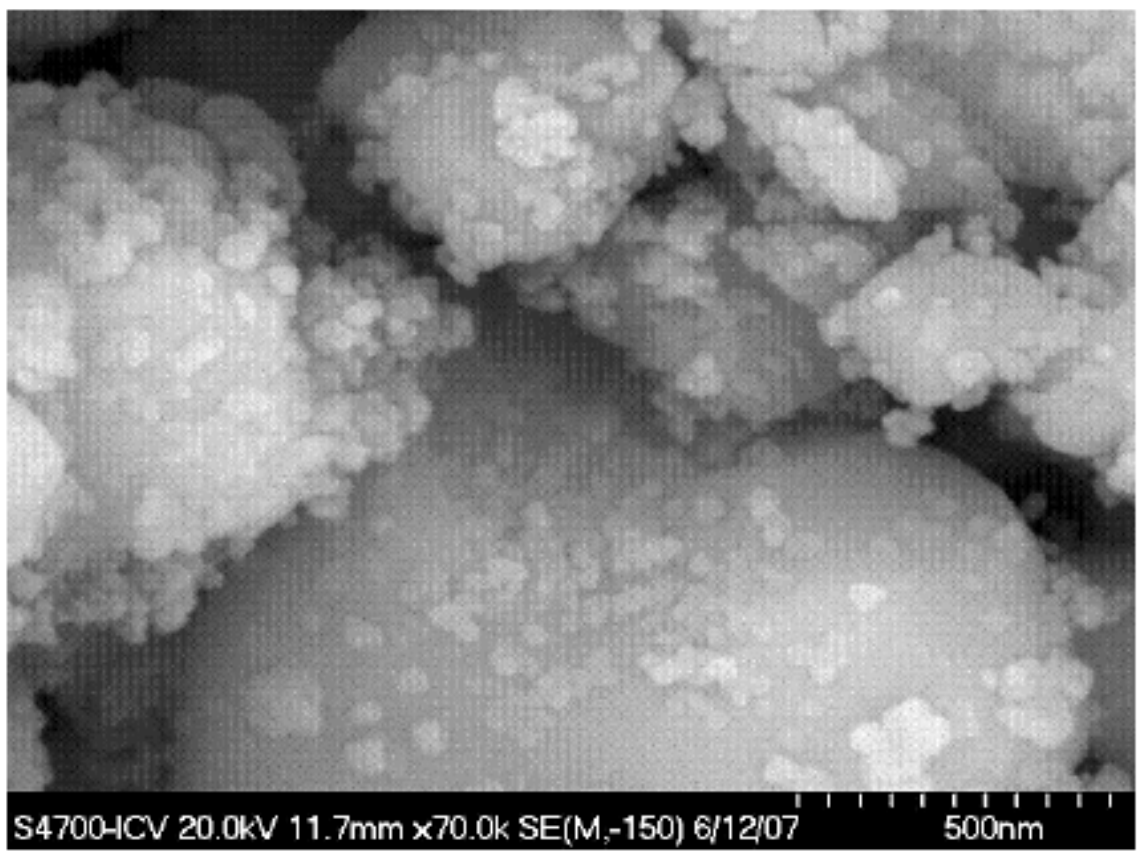

Figure 2. FE-SEM micrograph for the 5%$Co_3O_4$-95%ZnO sample after milling showing the $Co_3O_4$ particles dispersed over the surface of large ZnO ones.

## 3.2 Optical Absorption

Figure 3 shows pictures of the samples were clear differences in colour are observed. While ZnO and $Co_3O_4$ show white and black colour respectively, milled samples and those annealed at 500ºC and 600ºC resulted blue (a colour that can not be achieved by combination of black and white), confirming that there is a modification of electronic structure just after mixing. The sample annealed at 800ºC results greenish which is the characteristic colour of $ZnCo_2O_4$, in agreement with the XRD data (see figure 1).

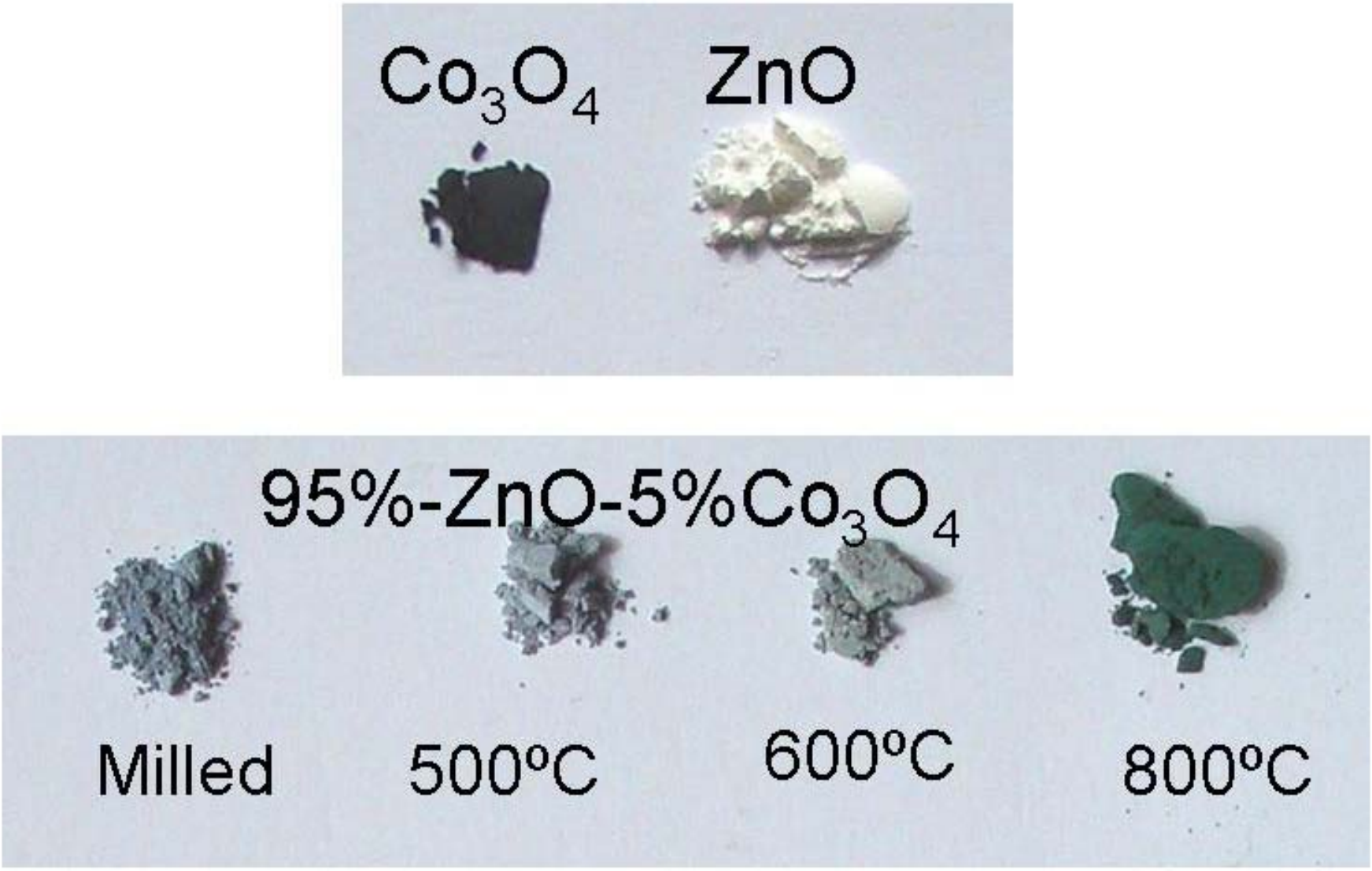

Figure 3. Pictures of (top) ZnO and $Co_3O_4$ powder and (bottom) Mixtures 5%Co3O4-95%ZnO after milling and annealing at different temperatures.

A more detailed analysis of these visual changes can be performed by means of optical spectroscopy. Figure 4a displays the optical reflectance spectra in the UV-Vis part of the spectrum. Pure ZnO spectrum exhibits the characteristic shape for semiconductors with a flat shape over the bandgap (around 400 nm) and a sharp decrease at the edge, while $Co_3O_4$ one shows a "surfing profile" with very small reflectance values. The samples containing 5% $Co_3O_4$ just milled and those annealed at 500º and 600º (i.e., the magnetic samples) shown a non flat profile spectrum with a maximum of the reflectance at 400 nm associated with the samples blue colour.

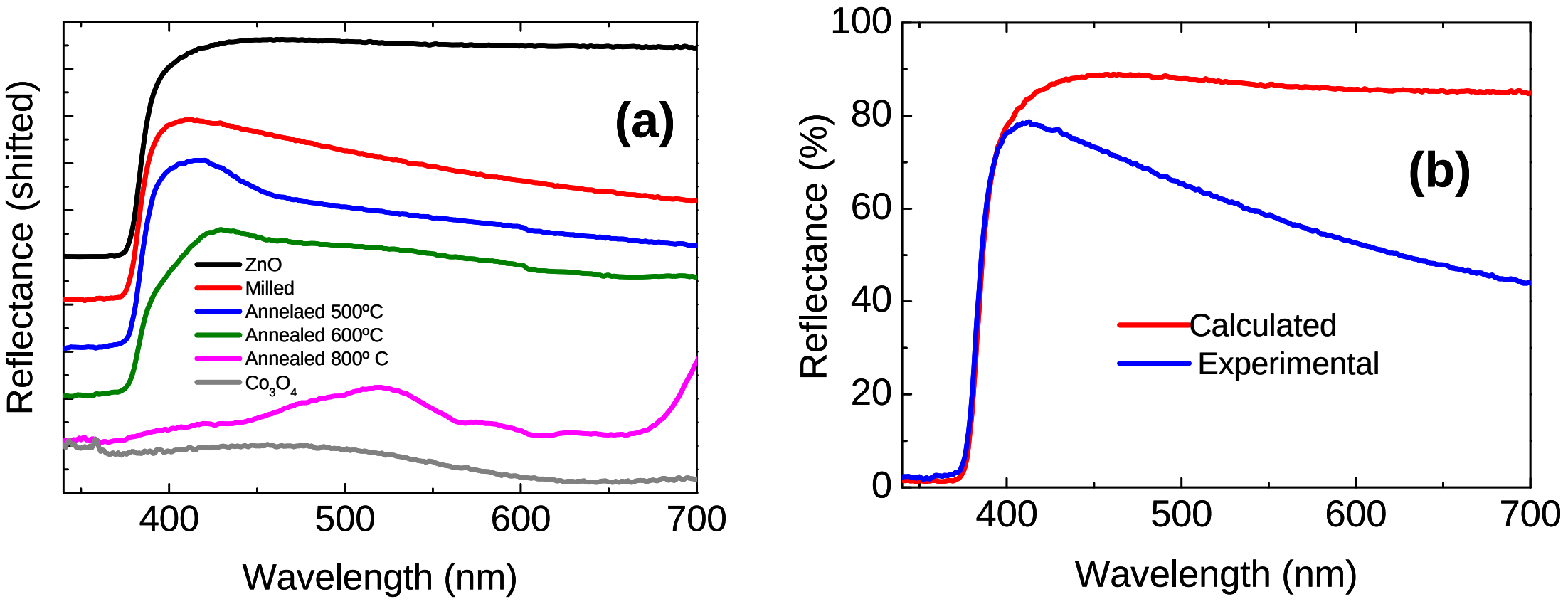

Figure 4. (a) Diffuse reflectance spectra of the samples (curves are shifted vertically for clarity) (b) Spectra for 5%$Co_3O_4$-95%ZnO experimentally measured for the milled samples and that calculated by linear combination of ZnO and $Co_3O_4$ spectra.

This maximum is due to the combination of the ZnO band-edge at lower wavelengths and a broad absorption band in the visible part of the spectrum. The sample annealed at 800ºC shown a complete different spectrum with strongly reduced reflectance and a maximum at about 500 nm responsible of the green colour.

Figure 4b presents the spectrum obtained by linear combination of 5%$Co_3O_4$ and 95% of ZnO and that corresponding to the milled sample. The evident differences between both spectra, confirms the reaction between ZnO and $Co_3O_4$ after milling, leading to a modification of the electronic structure that is the ultimate responsible of the material optical properties. This modification of the electronic configuration of ZnO (and consequently on that of $Co_3O_4$) just after milling seems to be in the origin of the observed FM. Actually, the three samples exhibiting RT FM show a very similar reflectance spectra profile, supporting this hypothesis.

Blue colour is characteristic of compounds containing $Co^{+2}$ in tetrahedral positions that arise two absorption peaks centred at 435 nm and 729 nm [14]. These peaks correspond to the ligand to metal ($p(O^{2-}) \rightarrow e_g(Co^{3+})$) and metal to metal ($t_{2g}(Co^{3+}) \rightarrow t_2(Co^{2+})$) charge transfer respectively. However the light blue coloration in our samples is fairly different to that characteristic of materials containing tetrahedral $Co^{2+}$ (that tends to be darker). Actually, we just find a very broad absorption band in the reflectance spectrum (figure 4) in contrast with the two double bands indicated above.

A broad absorption band in the Vis part of the spectrum and blue coloration similar to that reported here have been ascribed to the coexistence on $Co^{2+}$ in both tetrahedral and octahedral positions in different matrices [15,16]. $Co^{2+}$ in octahedral position produce absorption bands that overlaps those of the tetrahedral Cobalts, yielding to a broad absorption spectrum [17]. For $Co_3O_4$ tetrahedral positions are occupied by $Co^{+2}$ while $Co^{+3}$ are placed in octahedral ones [18]. Thus, there are no $Co^{+2}$ ions in octahedral positions. However, if $Co^{+3}$ ions in octahedral position are reduced to $Co^{+2}$ without any modification of the crystallographic structure, the light blue colour and a broad absorption band could be explained. Such a reduction could be promoted by the interaction of ZnO considering the strong electrostatic interaction between both oxides [12,13]. During milling process some $Co^{+2}$ in the surface of $Co_3O_4$ could be reduced keeping the crystal structure (as evidenced by XRD). This region should be small and close to the surface. A massive reduction of $Co^{+3}$ to $Co^{+2}$ will rend the crystallographic structure unstable and the oxide will transform into CoO. The absence of changes in the XRD patterns after mixing indicates that any modification due to the interaction between both oxides is limited to the surface region where both oxides are in contact.

However, there are many defects in ZnO that could also account for the observed blue colour [19]. In order to elucidate the possible reduction of Co ions we used an element specific technique as XAS.

### 3.3 X-ray Absorption Spectra

The X-ray absorption spectra measured at the Co K-edge are presented in figure 5a. The spectra for the samples with 5%$Co_3O_4$ milled and annealed up to 600ºC are very similar to that of pure $Co_3O_4$ in agreement with the XRD presented on figure 1. For the sample annealed at 800 ºC the spectrum is fairly different, due to the already mentioned formation of $Zn_{1-x}Co_xO$ solid solution.

Although standard XANES is a bulk spectroscopy, the resulting spectrum comes from the weighted contribution of every Co atom in the sample. In the case of nanoparticle systems, an enhanced contribution of atoms located at the surface is expected, rending XANE sensitive to surface modifications [20].

These spectra show a pre-edge feature (figure 5b) with a maximum at about 7707 eV This maximum corresponds to a forbidden quadrupolar transition 1s→3d, in which the 3d level presents a p character due to pd hybridization. However, after milling, a second weak peak appears at energies slightly larger than the main peak (around 7710 eV); this second peak is present for milled samples and annealed at 500ºC while it can not be observed for larger annealing T (i.e.; it is present in the samples exhibiting magnetic behaviour). This suggests a slight modification in the electronic structure of cobalt, which can impinge on the FM observed. In order to understand these modifications, a qualitative study by means of XANES calculations has been performed.

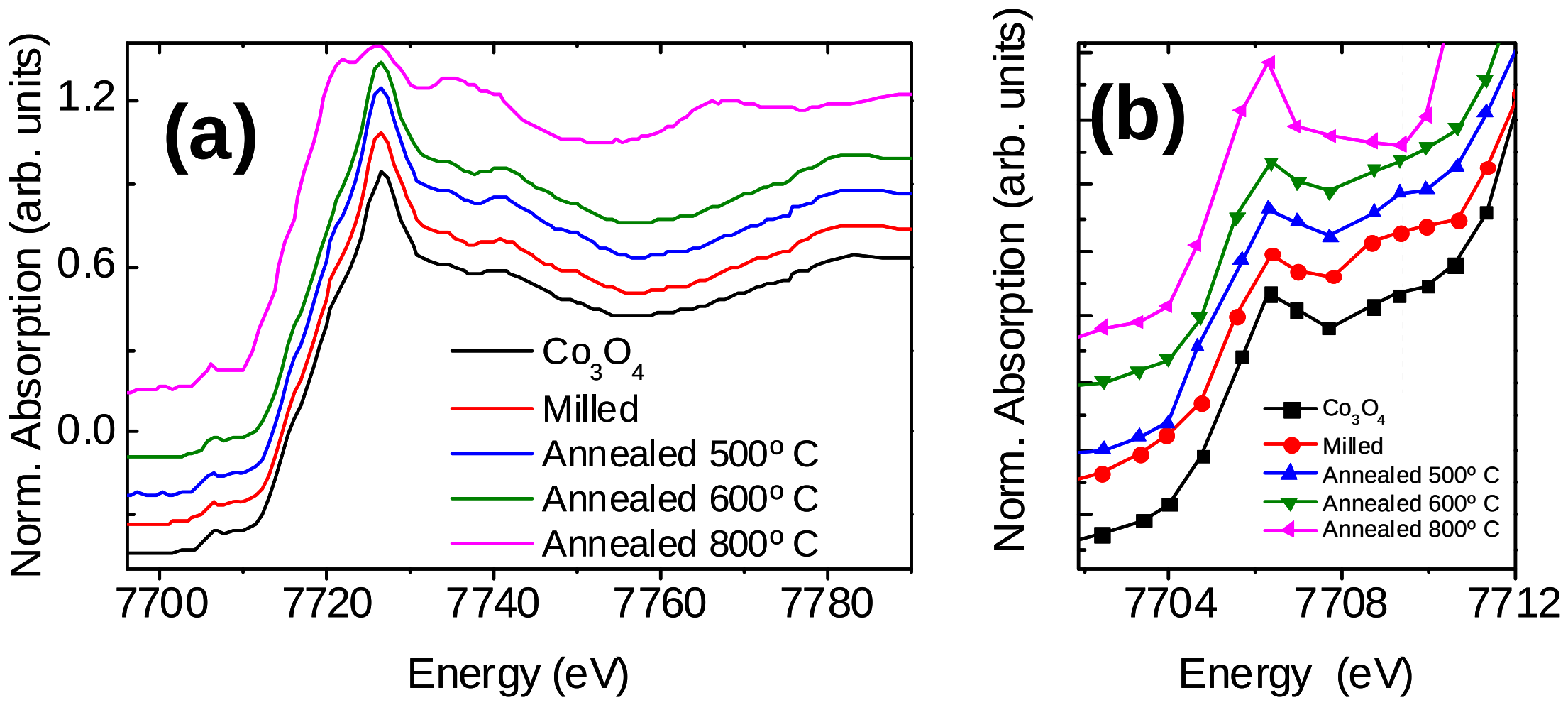

Figure 5. (a) XANES Co K-edge spectra of the samples; a pure $Co_3O_4$ XANES has been added as reference; (b) Detail of the XANES spectra at the pre-edge region. (Curves are shifted vertically for clarity)

XANES calculations have been performed within the real-space multiple-scattering formalism of the FEFF8 code. Self-consistent computations were achieved choosing a cluster of around 4 Å. Spherical atomic potentials have been proposed using the muffin-tin approximation, with default overlapping of spherical potentials of 15 %. Several exchange and correlation potentials have been tested. A complex Dirac-Hara (DH) electronic potential, in which we add the Hedin-Lundqvist (HL) imaginary part to the classical DH expression is found to perform a better reproducibility of both the position and the intensity of the experimental features. No extra broadening has been applied in the computations. Two clusters have been used to account for two different contributions according to the local structure around Co atoms: absorbing Co atom in octahedral position in a $Co_3O_4$ cluster (Co-oh core) and absorbing Co atom in octahedral position located at the interface between the $Co_3O_4$ and the ZnO (Co-oh interface).

Results of the calculations are shown in the figure 6. The calculated XANES spectra accounting for the two different environments (core and interface) show similarities in the position and the intensity of features located in the namely XANES region, above the white line. Visible differences are observed in the edge region. In the Co-oh interface calculated spectrum, the pre-edge region shows a double structure. The two pre-edge peaks appearing in the Co-oh interface cluster calculation evidence that the secondary feature appearing in the FM samples comes from the contribution of Co atoms at the interface (figure 5b). Furthermore, the secondary feature seems to present a remarkable intensity, in such a manner that even a very small contribution coming from the Co-oh interface atoms will lead to a small, but visible, enhancement of such feature in the experimental XANES spectrum. Moreover, the edge of the Co-oh interface XANES is found to be slightly shifted (≈2 eV) to lower energies, respect to the Co-oh core one. This feature, in agreement also with the decrease of intensity in the white line in the interface contribution, suggests a decrease in the oxidation state of the octahedral Co atoms at the surface, which supports the idea of the reduction from $Co^{3+}$ to $Co^{2+}$ in the octahedral Co atoms located at the interface regions.

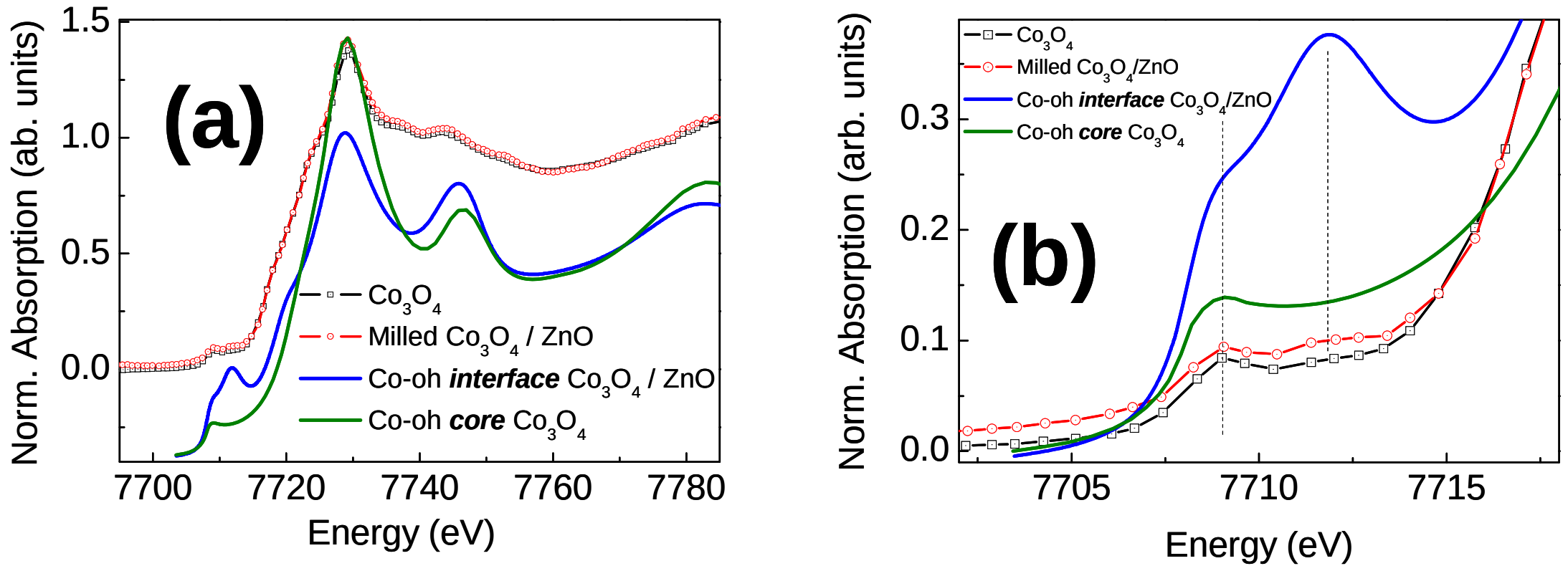

Figure 6. (a) Calculations of the contribution to the XANES of absorbing Co octahedral atoms located at the core of the $Co_3O_4$ particle (green) and the Co3O4/ZnO interface (blue), compared with the milled sample and the $Co_3O_4$ reference; (b) Detail of the calculated and experimental XANES spectra at the pre-edge region

While we could not perform a definitive experiment to confirm the reduction of $Co_3O_4$, it is worthy to note that this hypothesis allows explaining simultaneously the changes in the XAS spectra, optical spectra and magnetization curves. Thus, although not fully proved, the reduction of $Co^{+3}$ is strongly supported by the combination of experimental results. XAS spectra confirm the modification of the $Co_3O_4$ electronic structure; moreover the observation of magnetic features also points on this direction as ZnO is diamagnetic. The reduction of the $Co^{+3}$ produces an optical spectra (figure 4) very similar to that observed in the bibliography [15, 16] when a combination of $Co^{+2}$ in both tetrahedral and octahedral position are presented (a situation that can be achieved in $Co_3O_4$ by $Co^{+3}$ reduction); the shape of the experimental XANES pre-edge also fits the spectrum calculated for $Co_3O_4$ imposing a reduction of $Co^{+3}$.

A modification of the $Co_3O_4$ structure at the surface could also account for the observed modification but is not likely by several reasons. First, the soft milling performed to mix both oxides is not expected to induce structural changes and no modifications were found in the XRD patterns. Actually, after milling and annealing no structural changes are found up to annealing temperatures of 700ºC so it is rather unlikely that a structural change takes place during milling (note that no change in the particle size was found after milling according to SEM observations).

The pure $Co_3O_4$ milled in the same conditions showed no changes in the optical and XAS spectra nor in the magnetization curves; thus we can rule out a modification of the structure due to the mechanical milling process.

Therefore, the reduction $Co^{+3} \rightarrow Co^{+2}$ is likely and allows to explain the changes observed in the XAS and optical spectra (and as explained below, also in the magnetic properties)

The phenomenology described here, is fairly similar to that observed in mixtures of Anatase $TiO_2$-$Co_3O_4$ and can be interpreted in the same way [21]. $Co_3O_4$ possess the same spinel structure that $Fe_3O_4$, with $Co^{+2}$ in tetrahedral positions and $Co^{+3}$ in octahedral ones. In this kind of spinels, antiferromagnetic interactions between octahedral and tetrahedral cations are the dominant ones [18] leading to the well known ferrimagnetic character of ferrites (due to the different number of sites and magnetic moment of octahedral and tetrahedral positions) with high Curie Temperature ($T_C$); e.g. 858 K for $Fe_3O_4$ or 673 K for $NiCo_2O_4$. As figure 7 illustrates, the magnetic moment of Co ions depends on the oxidising state and also on the symmetry of the crystal field, that determines the splitting of the doublet $t_{2g}$ and triplet $e_g$ states. For $Co_3O_4$, the $Co^{+2}$ ions in tetrahedral position hold no magnetic moments [18]. Thus, there are no antiferromagnetic interactions between tetrahedral and octahedral Co ions. In this situation, the weaker antiferromagnetic interaction between $Co^{+3}$ in octahedral positions become the dominant one, arising the well known antiferromagnetic character of this oxide with low Neel temperature (40K).

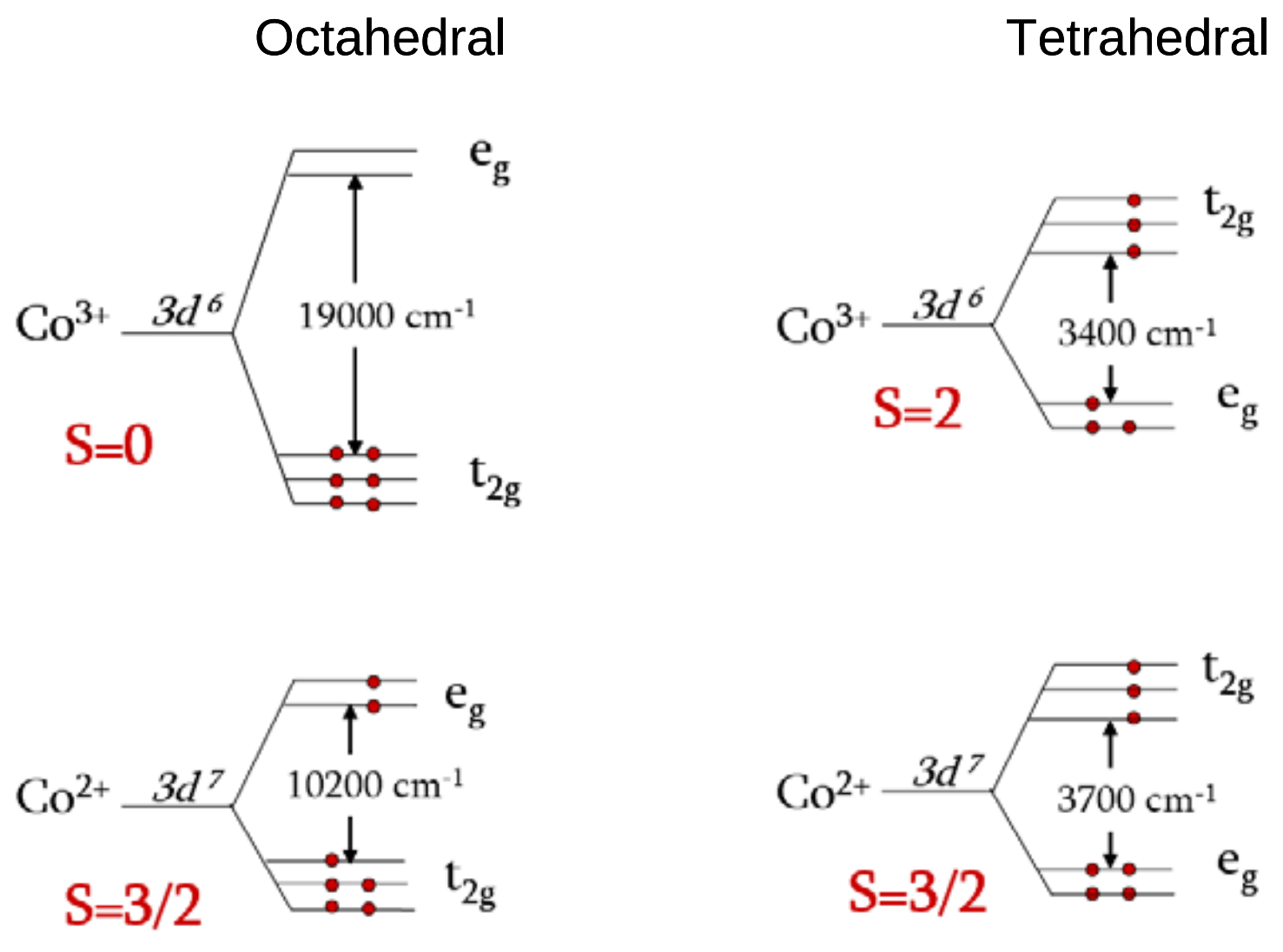

Figure 7. 3d band splitting for $Co^{+2}$ and $Co^{+3}$ ions in presence of an octahedral and tetrahedral crystal field.

If we assume that the reduction process suggested by XAS and optical experiments takes place without structural modifications, if there will be regions of the crystal close to the surface where $Co^{+3}$ is reduced to $Co^{+2}$: In this regions there will be Co ions holding a magnetic moment in both tetrahedral and octahedral positions. Then, the leading magnetic interaction will be antiferromagnetic between octahedral and tetrahedral positions. The situation will be very similar to that of $Fe_3O_4$ and the expected magnetic behaviour should be the same: ferrimagnetism with high $T_C$. Thus, we should expect a mixture of paramagnetism (from the grains core that remains unaffected) and small ferrimagnetic signal coming from the reduced surface regions. In this way, it is possible to explain the presence of room temperature ferromagnetism in the ZnO/ $Co_3O_4$ mixtures in basis of the well established theories of magnetism in oxides, and no new mechanism are required to account for the observed FM.

## 5. Conclusions

In summary, we demonstrated here that mixing and soft milling of ZnO and $Co_3O_4$ lead to a dispersion of the $Co_3O_4$ particles on the surface of large ZnO ones. Optical spectroscopy and XAS measurements strongly suggest that this interaction causes a surface reduction of $Co^{+3}$ in octahedral positions to $Co^{+2}$. In the $Co_3O_4$ surface where this reduction takes place, the structure and electronic configuration should similar to that for $Fe_3O_4$ explaining the observation ferromagnetic features (actually ferrimagnetic) in the magnetization curves with the superexchange interactions of oxides.

## 5. Acknowledgments

Lucas Perez and Manuel Plaza are acknowledged for the help with the magnetic measurements. M.S. Martín-González and J.L. Costa-Krämer are acknowledged for fruitful discussions. This work was supported by the Spanish Council for Scientific Research through the projects CSIC 2006-50F0122 and CSIC 2007-50I015 and Spanish Ministry of Science and Education through the projects MAT2007-66845-C02-01 and FIS-2008-06249. We acknowledge the European Synchrotron Radiation Facility for provision of synchrotron radiation facilities and we would like to thank the SpLine CRG beamline staff for assistance during X-Ray absorption experiments